\documentclass[prb, aps, twocolumn, amsmath, amssymb, superscriptaddress,noeprint]{revtex4-2}

\usepackage{bbold}
\usepackage{float}
\usepackage[dvips,final]{graphicx}
\usepackage{dcolumn}
\usepackage{bm}
\usepackage{textcomp}
\usepackage{amsmath,amssymb,amsfonts}
\usepackage{color}
\usepackage{hyperref} 
\hypersetup{colorlinks,linkcolor={blue},citecolor={blue},urlcolor={blue}}  
\usepackage{orcidlink}

\DeclareMathOperator*{\argmin}{\arg\!\min}

\begin{document}
\let\WriteBookmarks\relax
\def\floatpagepagefraction{1}
\def\textpagefraction{.001}

\title{Adaptive Time Stepping for the Two-Time Integro-Differential Kadanoff-Baym Equations}

\author{Thomas Blommel \orcidlink{0009-0008-7101-3534}}
\affiliation{Department of Physics, University of Michigan, Michigan 48109, USA}

\author{David J. Gardner \orcidlink{0000-0002-7993-8282}}
\affiliation{Center for Applied Scientific Computing, Lawrence Livermore National Laboratory, California 94550, USA}

\author{Carol S. Woodward \orcidlink{0000-0002-6502-8659}}
\affiliation{Center for Applied Scientific Computing, Lawrence Livermore National Laboratory, California 94550, USA}

\author{Emanuel Gull \orcidlink{0000-0002-6082-1260}}
\affiliation{Department of Physics, University of Michigan, Michigan 48109, USA}

\date{\today}
\begin{abstract}
The non-equilibrium Green's function gives access to one-body observables for quantum systems.  Of particular interest are quantities such as density, currents, and absorption spectra which are important for interpreting experimental results in quantum transport and spectroscopy.  We present an integration scheme for the Green's function's equations of motion, the Kadanoff-Baym equations (KBE), which is both adaptive in the time integrator step size and method order as well as the history integration order. We analyze the importance of solving the KBE self-consistently and show that adapting the order of history integral evaluation is important for obtaining accurate results.  To examine the efficiency of our method, we compare runtimes to a state of the art fixed time step integrator for several test systems and show an order of magnitude speedup at similar levels of accuracy.
\end{abstract}
\maketitle

\section{Introduction}\label{sec:Kanadoff Baym}

Quantum systems exposed to strong time-dependent perturbations, such as pulses or quenches, exhibit interesting and technologically useful phenomena that cannot be observed in equilibrium.  For example, the emergence of ultrafast spectroscopy has allowed for the probing of phenomena on the attosecond timescale \cite{Atto2009}, such that the quasiparticle dynamics and Higgs mode oscillations of high temperature superconductors  can be studied \cite{Giannetti2016}, and  retardation effects and photoabsorption peak splitting in atoms can be investigated \cite{Perfetto2015}.
Beyond simple limits, understanding this physics requires the numerical simulation of the time-dependence of the quantum system \cite{DMFT,Werner2010}.  In cases where both the quantum nature of the system and the time dependence of the perturbation are non-trivial, standard approaches, such as density matrix renormalization group (DMRG) \cite{DMRG, DMRG2}, time dependent density functional theory \cite{TDDFT,TDDFTBook}, and quantum Monte Carlo (QMC) \cite{iQMC,RTQMC}, are either restricted to short times, small systems sizes, or make approximations to the time-dependence that are difficult to justify in general.

Quantum field theory, in particular the Keldysh diagrammatic approach, presents an elegant theoretical formulation of time-dependent problems \cite{Kadanoff1962Book}.  Derivations of equations of motion and conservation laws are presented in \cite{SvL2013}, while the action and path-integral approaches are discussed in \cite{Kamenev_2011}.  
The equations of motion of this theory, known as the Kadanoff-Baym equations (KBE) \cite{SvL2013},  are a set of coupled integro-differential equations.  Numerically, reaching long enough final times for the propagation with controlled accuracy both in the integration of the equation of motion 
and the calculation of the quantum correlation effects, has proven to be challenging due to the cubic scaling of the integration \cite{nessi,BalzerBook}.

Substantial progress has been achieved in recent years using approximate methods.  The Generalized Kadanoff Baym Ansatz (GKBA) \cite{GKBA_original,Hermanns_2012} integrates an approximate form of the equations and has found increased usage recently due to the introduction of a linear scaling formulation \cite{G1_G2}.  The approximation has numerically been shown to neglect terms around two orders of magnitude smaller than those kept \cite{Reeves2023}.  The GKBA has been used to study a wide range of systems, including electron-boson coupling~\cite{EB_DynamicsGKBA}, exciton dynamics~\cite{Exciton_GKBA}, surface charge dynamics~\cite{Surface_GKBA}, and quantum transport~\cite{QT_GKBA}.  These calculations would benefit from benchmarking with the KBE due to observed differences between KBE and GKBA~\cite{GKBA_KBE_bench}. The memory truncation method \cite{Truncation} introduces a cutoff time, where history information older than the cutoff is discarded.  This has been shown to be  effective at decreasing the scaling of integration costs, and recovers dynamical mean-field theory dynamics well \cite{AcceleratedCollapse}.  Recently, extrapolation techniques have also been successful in learning dynamics from short time samples of data.  This has been accomplished for dynamics along the diagonals of the Green's function \cite{DMD1T,DMD2T}, which contain density information, as well as for dynamics along one time coordinate, which gives spectral information \cite{Extension}.

There has also been much progress in solving the full KBE without approximations.  The hierarchical compression scheme \cite{KayeComp2021} exploits the compressibility of the two-time Green's functions to lower the scaling of the computational cost and memory requirements.
Adaptive integration techniques have also been applied to the KBE \cite{meirinhos2022adaptive}, showing an ability to resolve dynamics on time scales separated by orders of magnitude.  All of these methods bring physically relevant time scales into reach and further bridges the gap over the ``valley of death'' between short-time transient dynamics and long-time steady-state physics. This paper revisits the approach of adaptive time integration for the KBE, presenting a methodology that allows for substantially higher precision on longer time scales than were previously accessible.   In order to gauge the efficiency of this adaptive integrator against previously developed fixed-step integrators, we will be focusing on the dynamics of the density matrix for several model systems.  This method is not limited to density information however, and physical observables such as energy, spectra, order parameters, and photoabsorption spectra are easily able to be extracted from the Green's function data.

The remainder of this paper is organized a follows.
In Section~\ref{Sec:KBE} we introduce the Kadanoff Baym equations. Section~\ref{Sec:Solution} presents our adaptive solution strategy for advancing the KBE in time and computing the history integrals.
We give an overview of the numerical implementation of these algorithms in Section~\ref{Sec:Implementation}. 
Numerical results for a simple model of the hydrogen molecule are given in Section~\ref{Sec:Simple_Results} and are followed by results from a Hubbard model in Section~\ref{Sec:Hubbard_Results}.
To study the inclusion of the thermal branch, we analyze a simple model of photoexcited superconductors in Section~\ref{Sec:thermal}. Concluding remarks and directions for future work are discussed in Section~\ref{Sec:Conclusions}.

\section{Kadanoff Baym Equations}\label{Sec:KBE}

The Kadanoff Baym equations \cite{SvL2013,Kadanoff1962Book} are a system of coupled integro-differential equations which describe the dynamics of non-equilibrium Green's functions for quantum systems.  The Green's function is a complex, matrix-valued function of two types of variables, real times are called $t$ and $t'$, while imaginary times, which are used to account for thermal effects, use $\tau$. 
The matrix dimension, $M$, of the Green's function corresponds to the number of orbitals.  The variables $t$ and $t'$ vary from 0 to $T_f$, the final time of interest in the calculation, while $\tau$ varies from 0 to $\beta$, which is the inverse of the system temperature.  These two types of arguments arise from the three-legged Keldysh contour, and the ``Keldysh'' components of the Green's function are defined based on the location of these arguments on the contour \cite{SvL2013}.  The lesser, $G^<(t,t')$, the greater, $G^>(t,t')$, the mixed, $G^\rceil(t,\tau)$, and the Matsubara, $G^M(\tau)$, components are needed to fully describe the quantum system. These functions are coupled to each other through the KBE and must be solved for simultaneously.  The lesser and greater functions share an anti-Hermitian symmetry given by
\begin{equation}
    G^\gtrless(t,t') = -G^\gtrless(t',t)^\dagger,
\end{equation}
where $\dagger$ denotes the conjugate transpose of the matrix. Because of this symmetry, we only need to calculate each component on one side of the $t=t'$ diagonal.  The Matsubara and mixed components are both anti-periodic with period $\beta$,
\begin{gather}
    G^M(\tau) = -G^M(\tau+\beta),\\
    G^\rceil(t,\tau) = -G^\rceil(t,\tau+\beta).
\end{gather}

The different components obey the boundary conditions
\begin{align}
    G^>(t,t) &= -i\mathbb{1} + G^<(t,t),\\
    G^<(0,t') &= -G^\rceil(t',0)^\dagger,\\
    G^\rceil(0,\tau) &= iG^M(-\tau).
    \label{Eq:BC}
\end{align}
It is useful to define linear combinations of the lesser and greater components of the Green's function, as they appear in the KBE
\begin{align}
    G^R(t,t') &= \Theta(t-t')[G^>(t,t') - G^<(t,t')],\\
    G^A(t,t') &= G^R(t',t)^\dagger,
\end{align}  
where $\Theta$ is the Heaviside function. 
The quantities $G^R$ and $G^A$ are known as the retarded and advanced components, respectively.
In the typical presentation of the KBE, there is a second mixed component that is usually defined as
\begin{equation}
    G^\lceil(\tau,t) = G^\rceil(t,\beta-\tau)^\dagger.
\end{equation}

Because the lesser and greater Green's functions are equations of two time arguments, there is a differential equation corresponding to each time argument, leading to a total of four equations
\begin{subequations} \label{Eq:KBE}
\begin{align}
    -i\partial_{t'} G^\gtrless(t,t') &= G^\gtrless(t,t')\epsilon(t') + \int_0^t d\bar{t} G^R(t,\bar{t})\Sigma^\gtrless(\bar{t},t') \nonumber\\& + \int_0^{t'}d\bar{t} G^\gtrless(t,\bar{t})\Sigma^A(\bar{t}, t')\nonumber\\& -i\int_0^\beta d\tau G^\rceil(t,\tau)\Sigma^\lceil(\tau,t'),\label{Eq:KBE1}\\
    i\partial_{t} G^\gtrless(t,t') &= \epsilon(t)G^\gtrless(t,t') + \int_0^t d\bar{t} \Sigma^R(t,\bar{t})G^\gtrless(\bar{t},t') \nonumber\\&+\int_0^{t'}d\bar{t} \Sigma^\gtrless(t,\bar{t})G^A(\bar{t}, t')\nonumber\\&-i\int_0^\beta d\tau \Sigma^\rceil(t,\tau)G^\lceil(\tau,t').
    \label{Eq:KBE2}
\end{align}
Furthermore, the mixed component is governed by the equation
\begin{align}
    i\partial_tG^\rceil(t,\tau) &= \epsilon(t)G^\rceil(t,\tau) + \int_0^td\bar{t}\Sigma^R(t,\bar{t})G^\rceil(\bar{t},\tau)\nonumber\\&+\int_0^\beta d\bar{\tau} \Sigma^\rceil(t,\bar{\tau})G^M(\bar{\tau}-\tau).
    \label{Eq:KBE3}
\end{align}
\end{subequations}
The function $G^M(\tau)$ must be supplied as an initial condition.  In the case where thermal expectation values are not of interest, the functions $G^M$ and $G^\rceil$ may be discarded.  This amounts to removing \eqref{Eq:KBE3} from the system of equations as well as dropping the $\tau$ integrals present in \eqref{Eq:KBE1} and \eqref{Eq:KBE2}.  In this case, the only initial condition that needs to be supplied is the initial value of the lesser Green's function $G^<(0,0)=i\rho(0)$.

The integrals appearing on the right hand side of the KBE are collectively called the history integrals.
The functions $\Sigma[G](t,t')$ appearing in \eqref{Eq:KBE} are known as the self-energies and arise from the interactions between the quantum particles the Green's function describes.  The self-energies are non-linear functionals of the Green's functions themselves, leading to self-consistency criteria for the KBE.  The exact form of the self-energy is dependant on the interacting Hamiltonian, as well as the diagrammatic approximation employed.  In our treatment, we shift the contribution of the Hartree-Fock self-energy away from $\Sigma$ and into the Fock matrix, $\epsilon$. In this work, we use the self-consistent second order perturbation theory (GF2) approximation \cite{Dahlen2005,Phillips2014,Holleboom1990} for the dynamical part of the self-energy 
\begin{align}
    \Sigma^\gtrless(t,t')_{ij} &= U_{ifde}U^\text{Ex}_{jabc}G_{fb}^\gtrless(t,t')G_{ea}^\gtrless(t,t')G_{cd}^\lessgtr(t',t),\nonumber\\
    \Sigma^\rceil(t,\tau)_{ij} &= U_{ifde}U^\text{Ex}_{jabc}G_{fb}^\rceil(t,\tau)G_{ea}^\rceil(t,\tau)G_{cd}^\lceil(\tau,t),\label{Eq:SE}\\
    U^\text{Ex}_{jabc} &= 2U_{jbac} - U_{jabc},\nonumber
\end{align}
where $U$ is known as the Coulomb tensor, and the indices denote orbitals.  The Fock matrix, $\epsilon(t)$ in \eqref{Eq:KBE2}, which describes single-particle interactions, 
also contains a dependence on the Green's function 
\begin{equation}
\begin{aligned}
    \epsilon_{ij}(t) &= h_{0,ij}(t) + \Sigma^{\text{HF}}_{ij}(t) \\ &= h_{0,ij}(t) + i(U_{iabj}-2U_{ijba}) G^<_{ba}(t,t), 
\end{aligned}
\label{Eq:eps}
\end{equation} where $h_0$ is the quadratic Hamiltonian of the system and $\Sigma^{\text{HF}}_{ij}(t)$ denotes the static Hartree-Fock self-energy. 
This non-linear dependence is typically not made explicit when the KBE are presented and is meant to be assumed. We will be following this convention for our presentation of the equations.  The self-energy obeys the same anti-Hermitian symmetry condition as the Green's function, and we store its Keldysh components in the same way.

It is possible to reformulate the KBE to be a system of one of the lesser/greater components and one of the retarded/advanced components. The NESSi package \cite{nessi} and recently developed integration schemes which leverage compression techniques \cite{KayeComp2021} use this formulation. 
In contrast to the formulation in terms of greater and lesser Green's functions, this formulation has the advantage of cheaper history integrals; the retarded equation has only one integral over the interval $[t',t]$.  Here, we choose the formulation in terms of the lesser and greater components because it is easier to obtain high order accurate integrals as discussed in Section~\ref{SSec:IW}. Because of the symmetry of the system, we opt to solve $G^<$ on and above the $t$-$t'$ diagonal and $G^>$ below the diagonal, meaning we use the lesser form of \eqref{Eq:KBE1} and the greater form of \eqref{Eq:KBE2}, so that two rather than four equations need to be solved.

\section{Solution Approach}
\label{Sec:Solution}

In this section we present adaptive methods for advancing the KBE in time and methods for computing the history integrals therein. 
Adaptive formulations for the KBE have previously been presented in \cite{meirinhos2022adaptive}. As we will elaborate below, the main differences between the formulation presented here and in \cite{meirinhos2022adaptive} are that we include the Matsubara branch, study the effects of self-consistency, and implement adaptive orders in the evaluation of history integrals.
We first recognize that \eqref{Eq:KBE} contains three (greater, lesser, and mixed) functions which need to be integrated alongside each other.
At first it may seem like the most efficient way of proceeding is to integrate the functions separately, which would allow the adaptive method to find optimal time steps for each function separately. 
This separation would be useful in the case where the dynamics of the different functions exist on largely different time scales.  Examining the history integrals present in the equations of motion, however, shows that all functions need to be known on the same time grid. 
This dependence means that if the functions are integrated separately, there will be large overheads and potential inaccuracies due to interpolating between grids.  
Furthermore, there is no reason to expect the functions to have large discrepancies in their dynamics, which is closely related to the energy landscape of the physical system being studied.

\begin{figure}[htbp!]
    \centering
    \includegraphics[width=0.5\textwidth]{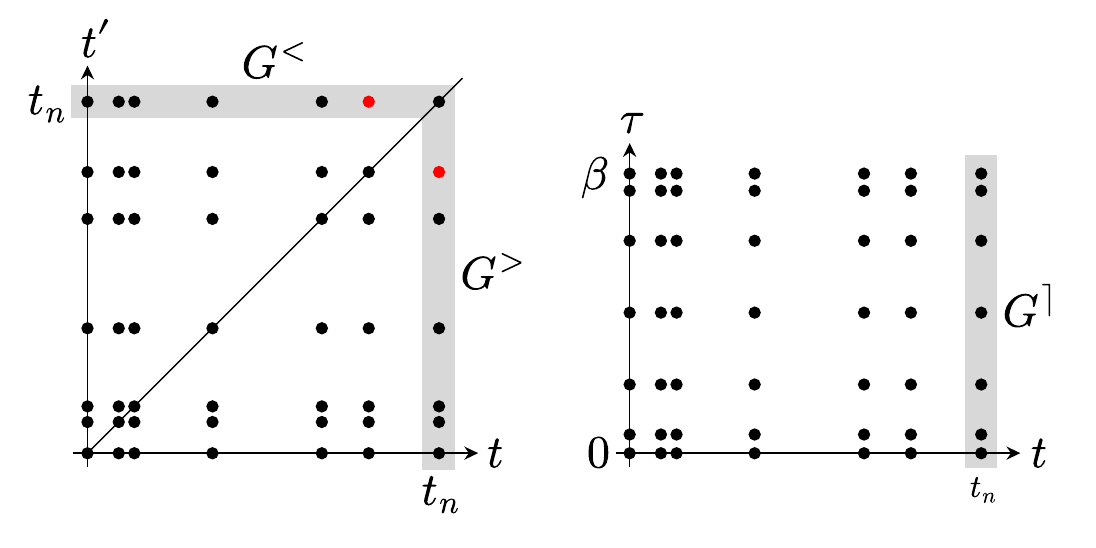}
    \caption{Integrating the KBE for $G^\gtrless$ amounts to filling the $(t,t')$ plane outwards from the origin.  We choose to represent the lesser (greater) function on the upper (lower) part of the plane. For the mixed component, the solution vector is enlarged by the number of points are on the $\tau$-axis.  The solution vector at time $t_n$ is contained within the shaded region.  The red elements within the solution vector indicate points that must have their history supplied manually in order to construct a Nordsieck array for adaptive time stepping. All other elements already have a history available.   
    (Figure adapted from one appearing in \cite{meirinhos2022adaptive}.)
    }
    \label{fig:fan}
\end{figure}

The integration of the Keldysh functions is therefore considered to take place on a grid with the same, non-uniform spacing in the $t$ and $t'$ directions.  For the mixed component, integration occurs in the $t$ direction, at points $\tau_i\in [0,\beta]$, with step sizes chosen to match those of the lesser/greater components.
This method of evolving the system with adaptive time integration methods was proposed and studied in \cite{meirinhos2022adaptive}, which we build upon here to study the importance of self-consistency on the KBE, analyze the effects of adapting the integration order of the history integrals, and the inclusion of thermal effects. 
As in \cite{meirinhos2022adaptive}, we ``stack'' the solution vectors and treat them as a system of differential equations in a single time variable, $\tilde{t}$, with the state vector $y(\Tilde{t})|_{t_n} = \{G^\rceil(\Tilde{t},\tau_j), G^>(\Tilde{t},t_i),  G^<(t_i, \Tilde{t}), G^<(\Tilde{t},\Tilde{t})\}$ for $0\leq t_i < t_n$ and $0\leq j < N_\tau$.  
This propagation scheme is illustrated in Figure~\ref{fig:fan}, where $y(\tilde{t})$ contains the points in the shaded grey region.

\subsection{Adaptive Time Stepping}
\label{SSec:ATS}

Considering the KBE as system in terms of the combined solution vector, $y(\Tilde{t})$, we can recast the problem in the form
\begin{equation} \label{Eq:ode}
    \dot{y} = f(\Tilde{t}, y) \quad y(\Tilde{t}_0) = y_0,
\end{equation}
where $\dot{y} = dy/d\Tilde{t}$, the function $f(\Tilde{t}, y)$ is the combined vector of right-hand side (RHS) functions from \eqref{Eq:KBE}, and $y_0$ is the initial condition.
At the start of a step from $\Tilde{t}_{n-1}$ to $\Tilde{t}_n$ the state vector of $N_{n-1}$ complex valued components is first increased to $N_{n}$ components with $N_n = M^2(2n+1+N_\tau)$ for $n \geq 0$ where the $M^2$ factor arises due to the $M\times M$ matrix nature of the Green's function, and there are $n+1$ and $n$ points for the lesser and greater components, respectively.
The reason for the mismatch in the number of points between lesser and greater components is that we use \eqref{Eq:BC} to avoid solving for $G^>$ on the diagonal.
The $N_\tau$ factor comes from the number of points that lie on the $\tau$-axis, and is only included when integrating a thermal branch.
Thus each step adds $2M^2$ new complex values to the state by introducing two new points originating from the diagonal in the temporal grid as illustrated by the red dots in Figure~\ref{fig:fan}. 

We advance \eqref{Eq:ode} by applying variable-step Adams methods of orders 1 to 12 as described in \cite{byrne1975polyalgorithm} (and summarized below) for the EPISODE code and similarly implemented in the CVODE package \cite{hindmarsh2005sundials}. A time step with step size $h_n = \Tilde{t}_{n} - \Tilde{t}_{n-1}$ using a method of order $q$ begins with the polynomial interpolant, $\pi_{n-1}$, satisfying the $q + 1$ conditions
\begin{equation} 
\label{Eq:predictor}
\begin{aligned}
    \pi_{n-1}(\Tilde{t}_{n-1}) &= y_{n-1}, \\
    \dot{\pi}_{n-1}(\Tilde{t}_{n-j}) &= \dot{y}_{n-j}, \quad j = 1, 2, \dots, q, 
\end{aligned}
\end{equation}
where $y_n$ is the numerical approximation of $y(\Tilde{t}_n)$ and $\dot{y}_n$ is the approximation of $\dot{y}(\Tilde{t}_n) = f(\Tilde{t}_n, y(\Tilde{t}_n))$. The solution at time $\Tilde{t}_n$ is obtained by constructing the polynomial interpolant, $\pi_{n}$, satisfying the $q + 2$ conditions
\begin{equation} 
\label{Eq:corrector}
\begin{aligned}
    \pi_{n}(\Tilde{t}_{n}) &= y_{n}, \\ \pi_{n-1}(\Tilde{t}_{n-1}) &= y_{n-1}, \\ \dot{\pi}_{n-1}(\Tilde{t}_{n-j}) &= \dot{y}_{n-j}, \quad j = 0, 1, \dots, q - 1.
\end{aligned}
\end{equation}
The solution history is represented using the Nordsieck array \cite{nordsieck1962numerical} given by 
\begin{equation} \label{Eq:Nordsieck}
    z_{n-1} = [\, y_{n-1},\; h_n \dot{y}_{n-1},\; h_n^2 \ddot{y}_{n-1} / 2,\; \ldots,\; h_n^q y_{n-1}^{(q)} / q! \,],    
\end{equation}
where the derivatives $y_{n-1}^{(j)}$ are approximated by $\pi_{n-1}^{(j)}(\Tilde{t}_{n-1})$. 
With this formulation, computing the time step is now a problem of constructing $z_n$ from $z_{n-1}$. 
This computation is done by forming the predicted array, $z_{n(0)}$, with columns $h_n^{(j)} \pi_{n-1}^{(j)}(\Tilde{t}_n) / j!$ for $j = 0, 1, \ldots, q$, and computing the coefficients $l = [l_0, l_1, \ldots, l_q]$ (see \cite{byrne1975polyalgorithm} for details) such that
\begin{equation} \label{Eq:z_update}
    z_n = z_{n(0)} + e_n l,
\end{equation}
where $e_n$ is the correction to the predicted solution i.e.,
\begin{equation} \label{Eq:correction}
   e_n = y_n - y_{n(0)} = \pi_{n}(\Tilde{t}_n) - \pi_{n-1}(\Tilde{t}_n).
\end{equation}
From the second column (index 1) of \eqref{Eq:z_update}, we obtain the nonlinear system
\begin{equation} \label{Eq:nls}
    F(y_n) \equiv y_n - \gamma f(\Tilde{t}_n, y_n) - a_n = 0,
\end{equation}
for computing $y_n$ where $\gamma = h_n / l_1$ and $a_n = y_{n(0)} - \gamma \dot{y}_{n(0)}$.

For non-stiff systems, \eqref{Eq:nls} is typically solved using a fixed point iteration. For this work we use the convergence criteria $R \| y_{n(m)} - y_{n(m-1)} \|_{\textsc{wrms}} < 0.1  / |C|$ to halt the iteration where $R$ is an estimate of the convergence rate, $y_{n(m)}$ is the $m$-th iterate, and $C$ is the leading coefficient of the local error estimate for the method (see \cite{hindmarsh2005sundials} for details). The weighted root-mean-square (\textsc{WRMS}) norm is defined as
\begin{equation} \label{Eq:wrms_norm}
    \|v\|_{\textsc{wrms}} = \left( \frac{1}{N} \sum_{j=1}^N \left(v_j\,w_j\right)^2\right)^{1/2},
\end{equation}
where $N$ is the length of the vector, $v$,  and the weights are defined by the relative and absolute temporal error tolerances (rtol and atol, respectively) as well as the most recent solution,
\begin{equation}
  \label{Eq:ewt}
  w_j = \left(\text{rtol}\, |y_{n-1,j}| + \text{atol}\right)^{-1}.
\end{equation}
The aim of this criteria is to ensure the iteration error is small relative to the temporal error \cite{hindmarsh2005sundials}. 

As explained in \cite{byrne1975polyalgorithm}, after successfully solving \eqref{Eq:nls} for $y_n$, the correction $e_n$ in \eqref{Eq:correction} can be utilized to obtain an estimate of the local truncation error (LTE) in the step. If the step fails the error test, i.e., $\|\text{LTE}\|_{\textsc{wrms}} > 1$, it is rejected, and the LTE is used to estimate the step size for a new step attempt. On a successful error test, we can use $e_n$ to estimate the LTE for a method of order $q - 1$ and, by combining $e_n$ and $e_{n-1}$, we can also estimate the LTE at order $q + 1$ (see \cite{byrne1975polyalgorithm} for details). These error estimates are used to select the method order that will maximize the next step size. Changes in step size for the current order are considered each time step while a change in order is only considered after $q + 1$ steps at order $q$. Safeguards are added to prevent too frequent changes in step size or order which can impact the stability of the method (see \cite{hindmarsh2005sundials} and the references therein for details).

As noted above, two additional points are added to the temporal grid at the start of each time step, and this addition increases the state vector length. Thus, after completing a step from $\Tilde{t}_{n-1}$ to $\Tilde{t}_{n}$, we must resize the solution vector for the state size at $\Tilde{t}_{n+1}$ and reconstruct the RHS history of $G^<(t_{n}, \Tilde{t}_{n+1})$ and $G^>(\Tilde{t}_{n+1}, t_{n})$ needed to create a new predictor polynomial \eqref{Eq:predictor}.
Once the necessary history is computed, the first column (index $0$) of the Nordsieck array is just the resized solution vector for the current time, and remaining columns can be filled using $\dot{y}_{n-j}$ for $j = 0, 1, \dots, q - 1$ to construct the interpolating polynomial and compute its derivatives.
Additionally, if an order change could occur in the next step, we must compute a new, resized correction vector, $e_{n-1}$, to estimate the LTE at order $q+1$ in the next step. 
This calculation is achieved by computing a new, resized prediction, $y_{n(0)}$, 
for the just-completed step and subtracting it from the just-computed solution, $y_n$.

\subsection{Adaptive History Integration}
\label{SSec:IW}

The lesser, greater, and mixed components of \eqref{Eq:KBE} require evaluating integrals of the form 
\begin{equation}
    I = \int_0^{t_n}d\bar{t} g(\bar{t}),
\end{equation}
where integrands are products of $\Sigma^{\gtrless,\rceil}$ and $G^{\gtrless,\rceil}$, which change at each time step as well as each fixed point iteration.  
Therefore, it is important that we are capable of computing highly accurate integrals with weights which do not depend on the function values and only on the past times.  The time points where function values are stored correspond to the times reached by the adaptive integrator at the completion of each step.  
Values are stored at the computed adaptive step time points rather than an equidistant time grid to avoid interpolating the history and potentially under-resolving or over-resolving certain domains, destroying any advantages of adaptivity. 

We accomplish this integration by building a weight matrix, $\omega_{ij}$, over the course of the evolution.
We begin the process of calculating these weights by splitting the integral into $I = \int_0^{t_{n-1}}d\bar{t} g(\bar{t}) + \int_{t_{n-1}}^{t_{n}}d\bar{t} g(\bar{t})$ and assuming that we know how to integrate functions on the range $[0,t_{n-1}]$, which has already been accomplished in the previous time step.  To perform the remaining integral, we explicitly integrate the $k^\text{th}$ order polynomial interpolant, $g(\bar{t})\approx\sum_{i=0}^k b_i(\bar{t}-t_{n-k})^i$, which is constructed using function values, $g(t_l)$, for $n-k\leq l\leq n$.  This process gives us
\begin{equation}
\begin{aligned}
    \int_{t_{n-1}}^{t_{n}}d\bar{t}\, g(\bar{t}) \approx&
    \sum_{i=0}^k b_i \int_{t_{n-1}}^{t_{n}}d\bar{t} \hspace{1.5pt} (\bar{t}-t_{n-k})^i, \\=&
    \sum_{i,j=0}^k V^{-1}_{ij} g_{n-k+j} T_i, \\=&
    \sum_{j=0}^k \omega^{(k)}_{j} g_{n-k+j},
    \label{Eq:weights}
\end{aligned}
\end{equation}
where we solve the linear system, $\omega^{(k)}_{i} = V^{-1}_{ij}T_i$, with $V$ being the Vandermonde matrix and $T_i = \int_{t_{n-1}}^{t_{n}}d\bar{t} \hspace{1.5pt} (\bar{t}-t_{n-k})^i$.  Note that the integration weights do not depend on the function values, only on the times at which the function is sampled.  In order to solve the Vandermonde system, which can be ill-conditioned, we use a fully pivoted LU decomposition, implemented in the Eigen library \cite{eigenweb} which, in our case, gives as accurate a solution as specialized Vandermonde inversion algorithms \cite{EISINBERG2006,ORUC2000}. 

What remains is to choose an optimal order, $k_\text{opt}$.  
If this order is chosen too small, the interpolation will be inaccurate, and if it is chosen too large, the interpolant may be a poor fit for the true function.  We choose this order by integrating a test function  $I^{(k)}_\text{Test} = \int_{t_{n-1}}^{t_{n}} d\bar{t}\,u(\bar{t})$ using all interpolation orders up to a maximal order, $K$, and taking the order that minimizes the difference from the result with the next lower order.  
The test function that we choose is 
\begin{equation}
    u(\bar{t}) = G^R(t_n,\bar{t})\Sigma^<(\bar{t},t_n) + G^<(t_n,\bar{t})\Sigma^A(\bar{t},t_n),
\end{equation}
which is the integral appearing in \eqref{Eq:KBE1}, the equation for $G^<(t_n,t_n)$.  We choose this component of the solution as it is of large importance due to its appearance in the Hartree-Fock self-energy in \eqref{Eq:eps}.  The optimal order, $k_{\text{opt}}$, is then chosen as 
\begin{equation}
    k_{\text{opt}} = \argmin_k{I^{(k)}_\text{Test}}-{I^{(k-1)}_\text{Test}}.
\end{equation}
This procedure of finding an optimal order does not incur large costs relative to the costs of computing the integrals required for each time step.  The test integrals are only computed over a very short interval and only one integral is done, as opposed to the $4n+2$ integrals required for an entire time step.  We have observed that this procedure is around 1\% of the cost of the total computation.  This procedure is repeated at every iteration of the fixed point solver, so as to ensure accuracy is not degraded due to changes between iterations.
Once the optimal order is found, we have a new row of integration weights
\begin{equation}
    \omega_{n,i}=
    \begin{cases}
        \omega_{n-1,i} \hspace{60pt} & \text{if} \hspace{10pt} i < n-k_\text{opt}\\
        \omega_{n-1,i} + \omega^{(k_\text{opt})}_{n-k_\text{opt}+i} \hspace{13pt} & \text{otherwise}
    \end{cases}.
\end{equation}
Note that $\omega_{n,i}=\omega_{n-1,i}$ is guaranteed for $i<n-K$, meaning the full $\omega$ matrix does not need to be stored.

\section{Implementation}
\label{Sec:Implementation}

The adaptive code used in this work is built upon the NEDyson package~\cite{NEdyson}, which originally used only a fixed step size and order time integration approach based on Backward Differentiation Formula (BDF) methods.
NEDyson can be broken into three major components that interface with each other to solve the KBE. 
First is the data storage class which leverages the anti-Hermitian properties of the Green's function and the self-energy by only storing values above the diagonal for the greater components and below the diagonal for the lesser components.
Next, is the module which evaluates $\Sigma^{\gtrless,\rceil}$ from the current value of $G^{\gtrless,\rceil}$ which is implemented to provide for easy changes in the functional dependence of $\Sigma^{\gtrless,\rceil}$ on $G^{\gtrless,\rceil}$ allowing a wide range of physical systems to be studied.
Last, is the module that evaluates the history integrals given $G^{\gtrless,\rceil}$ and $\Sigma^{\gtrless,\rceil}$.
In the fixed step case, these integrals are computed using the polynomial interpolation in \eqref{Eq:weights}, although with a fixed interpolation order, meaning all weights are able to be computed at the start.
For calculations which have a thermal branch, we represent functions of $\tau$ using the Discrete Lehmann Representation (DLR), which is implemented in the DLR package \cite{KayeDLR}.  This basis set is known to be highly compact when representing functions on the thermal branch.  The DLR package provides the points, $\tau_j$, $0\leq j < N_\tau$, that we sample from the thermal branch as well as tensors and matrices needed to perform the convolution in \eqref{Eq:KBE3} and integral in \eqref{Eq:KBE1} and \eqref{Eq:KBE2}, respectively.  

To apply the adaptive Adams methods discussed in Section~\ref{SSec:ATS} within NEDyson, we utilize the CVODE package from the SUNDIALS library \cite{hindmarsh2005sundials, gardner2022sundials} of time integrators and nonlinear solvers.
CVODE is a descendent of the EPISODE code described in \cite{byrne1975polyalgorithm} and similarly provides implementations of adaptive order and step size linear multistep methods for systems in the general form \eqref{Eq:ode} including variable coefficient Adams methods.
As with all SUNDIALS packages, CVODE is built on a shared core infrastructure including abstract interfaces for vectors, matrices, and algebraic solvers that encapsulate the methods in SUNDIALS from the data structures and parallelism employed by a specific class implementation.
Applications using SUNDIALS may supply their own class implementations or utilize any of the native versions included with SUNDIALS targeting a range of parallel computing paradigms \cite{balosEnablingGPUAccelerated2021}.
The problem defining function, $f(\tilde{t}, y)$ in \eqref{Eq:ode}, is provided to CVODE as a function pointer and is called as needed from within CVODE. Thus, to utilize Adams methods paired with a fixed point iterative solver from CVODE in NEDyson, we need a vector implementation to enable operating on the complex-valued state data and to define a function for computing the RHS of \eqref{Eq:KBE}. Additionally, to resize CVODE as the solution grows, we need to store the necessary state and derivative history to build the Nordsieck array as outlined in Section~\ref{SSec:ATS} and discussed further below.

When evolving the KBE, CVODE only needs to store and operate on data at the leading front in Figure~\ref{fig:fan}. However, NEDyson needs access to the entire history of $G^{\gtrless,\rceil}$ and $\Sigma^{\gtrless,\rceil}$, when computing the history integrals within calls to $f(\tilde{t}, y)$ from CVODE.
Thus, NEDyson maintains the tensors, $G_{tt'ij}$ and $G_{t\tau ij}$, for $G^{\gtrless,\rceil}$ (and corresponding tensors for $\Sigma^{\gtrless,\rceil}$) to store the values at all points reached by CVODE.
Tensor components where $t>t'$ store $G^>$, and components where $t\leq t'$ store $G^<$ (and similarly for $\Sigma^\gtrless$).
Access to these tensors within callbacks to $f(\tilde{t}, y)$ is achieved through a \texttt{void*} ``user data'' pointer given to CVODE and then passed back to NEDyson as an input to the RHS function called from CVODE. In this case, the pointer is to a structure containing pointers to instances of the NEDyson storage classes. 
When providing state data to CVODE, either at initialization or when resizing the integrator, the relevant entries of $G_{tt'ij}$ are copied into a SUNDIALS serial vector.
The state consists of $M^2(2n+1+N_\tau)$ complex values.
However, the SUNDIALS vector classes target real-valued problems. 
Thus, SUNDIALS vectors of $2M^2(2n+1+N_\tau)$ real entries are allocated and cast between real and complex arrays in order to copy data. 
The C and C++ standards ensure the representation and alignment of the arrays are compatible.
Additionally, we override two of the native SUNDIALS vector operations to compensate for storing the complex values as a vector of $2M^2(2n+1+N_\tau)$ real entries.
The function for computing the absolute value of the vector entries, used in \eqref{Eq:ewt}, is replaced by a function to compute the magnitude of the corresponding complex numbers and the WRMS norm function is modified to divide by $N/2$ rather than $N$ in \eqref{Eq:wrms_norm}.
Each time CVODE calls the RHS function, data must similarly be copied from the input SUNDIALS vector for the approximation of $y$ to the NEDyson storage class, so that NEDyson may evaluate $f(\tilde{t},y)$.
Once $G^{\gtrless,\rceil}$ is updated with the latest state approximation, the new self-energy, $\Sigma^{\gtrless,\rceil}$, is computed using native NEDyson functions, and the history integrals are evaluated as described in Section \ref{SSec:IW}.
These newly computed RHS values are saved directly into the RHS output vector provided by CVODE, and control over the integration returns to CVODE. 

Once CVODE successfully completes the step from $t_{n-1}$ to $t_n$, control returns back to the NEDyson driver loop where the solution vector is placed into the $G^{\gtrless,\rceil}$ class and the self-energy is evaluated one last time in order to be consistent with the new solution.
To begin the next step from $t_n$ to $t_{n+1}$, NEDyson must provide CVODE with information about the history of $G^{\gtrless,\rceil}$ and $f$ in order to build a predictor polynomial \eqref{Eq:predictor} for the new state size as discussed in Section~\ref{SSec:ATS}.
NEDyson copies values from the $G^{\gtrless,\rceil}$ storage class into SUNDIALS vectors for the two most recent solutions, $y_n$ and $y_{n-1}$, and evaluates $f(\tilde{t}, y)$ at the newly created points to fill $q + 1$ SUNDIALS vectors with RHS data, $\dot{y}_{n - j}$ for $j = 0, 1, \ldots, q$, where $q$ is the method order of the just completed step.
Most of the RHS values have been previously computed and saved to avoid repeated expensive history integrals.
This history reconstruction only amounts to the evaluation of $2 (q + 1)$ history integrals, which is small compared to the $2n+1$ required for each evaluation of $f(\tilde{t}, y)$.
This construction process requires the evaluation of the lesser (greater) version of \eqref{Eq:KBE} for $t>t'$ $(t'>t)$, which does not pose any issues as the equations are valid on both sides of the diagonal and there are no discontinuities.
Only $y_n$ and $\dot{y}_{n - j}$ with $j = 0, \ldots, q-1$ (or $q$ if the order is increased in the next step) are needed to build \eqref{Eq:predictor}.
The additional state vector and RHS vector are necessary to compute a new correction vector \eqref{Eq:correction} for the just completed step in order to estimate the error at the next higher order in the upcoming step.
Once all the history data is arranged by NEDyson, its passed to CVODE which internally constructs the new Nordsieck array using the current solution and a Newton interpolating polynomial for the derivative data.
Using the Newton form of the polynomial is advantageous as the higher order derivatives in the Nordsieck array can easily be computed recursively as needed from the Newton interpolant.

\section{Simple Ramp Results}
\label{Sec:Simple_Results}

In order to test the applicability of adaptivity to the KBE, we first test on a very simple model of molecular hydrogen.  We will be using the sto-6g basis set \cite{Hehre1969}, which leads to Green's functions that are $2\times2$ matrices.  
The quantities $h_{ij}$ and $U_{ijkl}$ in \eqref{Eq:SE} and \eqref{Eq:eps} are computed using the quantum chemistry package pyscf \cite{pyscf2020}.  We use the atomic unit system in these calculations where $h_{ij}$ and $U_{ijkl}$ are in Hartrees, [Ha], and all times are in inverse Hartrees, $\text{[Ha]}^{-1}$.  When exciting the system, we apply a rapid ramp on one orbital of the quadratic Hamiltonian~\cite{nessi,Reeves2023,quad_couple1} which is then ramped back down to its original value. The ramped quadratic Hamiltonian is given by
\begin{align}
    h_{ij} \rightarrow h_{ij} + \delta_{i0}\delta_{j0} \sum_{k=1}^n a_k (\tanh{[10(t - t_i)]} + 1),
    \label{Eq:ramp}
\end{align}
where $n = 4$, $a_1=a_2=-a_3=-a_4=1 \text{Ha}$, $t_1=1$, $t_2=3$, $t_3=10$, and $t_4=12$.  Each individual ramp has approximately a 10as width which is shorter than, but on the order of the current state-of-the-art pulse widths~\cite{atto}.  Despite this, our integrator is able to easily capture the dynamics of the system around this excitation.

This ramping should have several effects on the time step sizes in the calculation.  First, the rapid ramping function will need to be resolved well, leading to very small time steps taken in the vicinity of the ramps.  Second, the Green's functions will oscillate much quicker in the presence of the ramp, leading to smaller time steps even when the ramp function is steady.  The ramp will eventually return to zero, and the system should return to a steady-state solution. 
At steady state, the Green's functions oscillate much more slowly, and the step sizes can increase significantly from the excited state values.

To evaluate the accuracy of the adaptive method, we compare the results against a previously validated reference solution produced with the constant time step code using a sixth order BDF method, as in \cite{nessi}.
We ensure that the reference solution is self-converged at a timestep size of $h=0.01$ 
with differences between the reference solution and a solution with step size of $h=0.015$ which are an order of magnitude less than differences between the reference solution and solution of the adaptive code with the tightest tolerances.
We investigate the convergence of two quantities: the density matrix at the end of the dynamics introduced by the ramps and the slope of the density matrix in the steady-state regime, which should be zero.  When setting the tolerances, we choose the absolute tolerance to be $10^{-10}$. This tolerance is very small because the solutions are highly oscillatory and cross zero often.  We then choose a variety of relative tolerances that give a range of several digits of accuracy in the density matrix. 
All timings were performed on a single core 4th generation AMD Epyc CPU.

\subsection{History Integral Evaluation}

\begin{figure}[htbp!]
    \centering
    \includegraphics[width=0.23\textwidth]{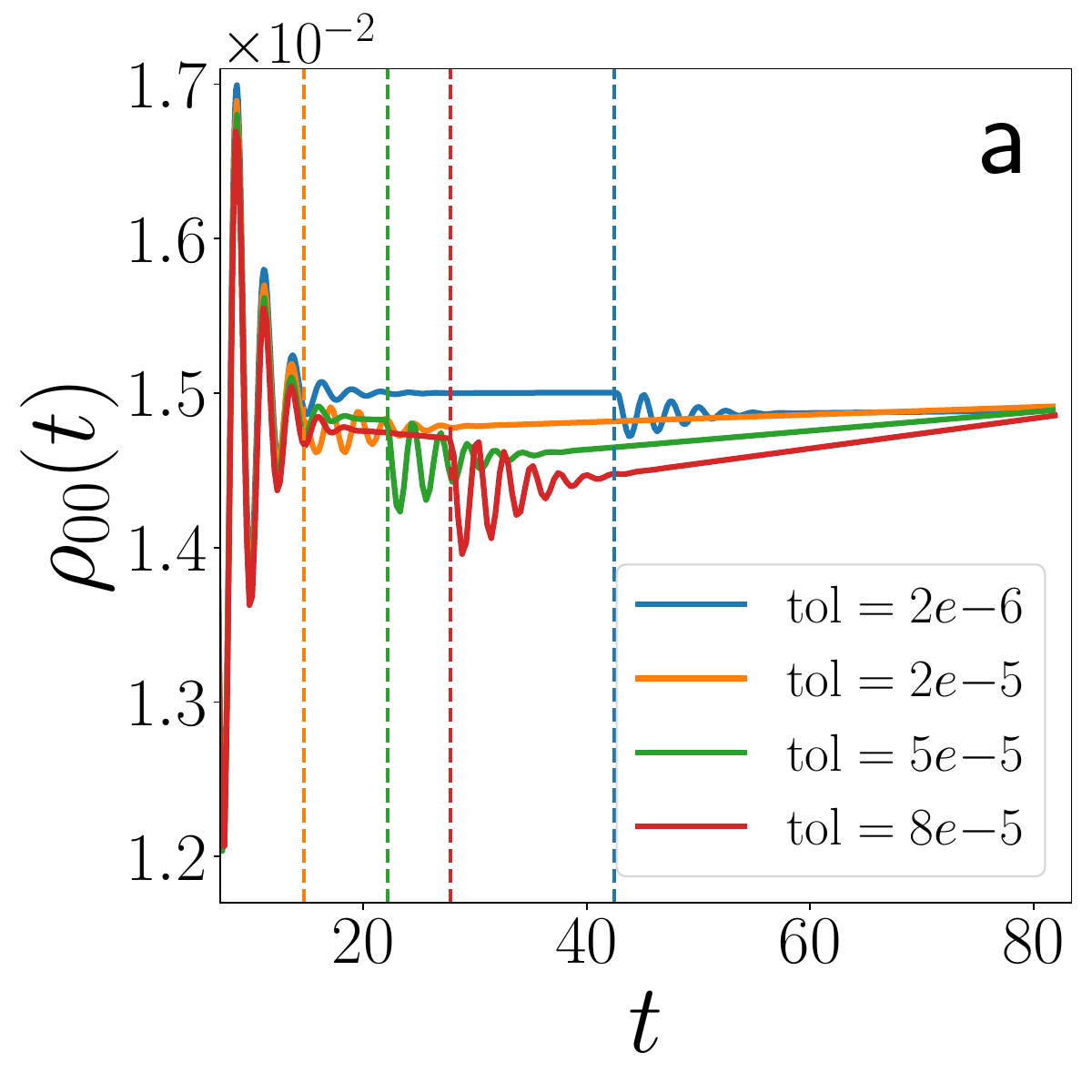}
    \includegraphics[width=0.23\textwidth]{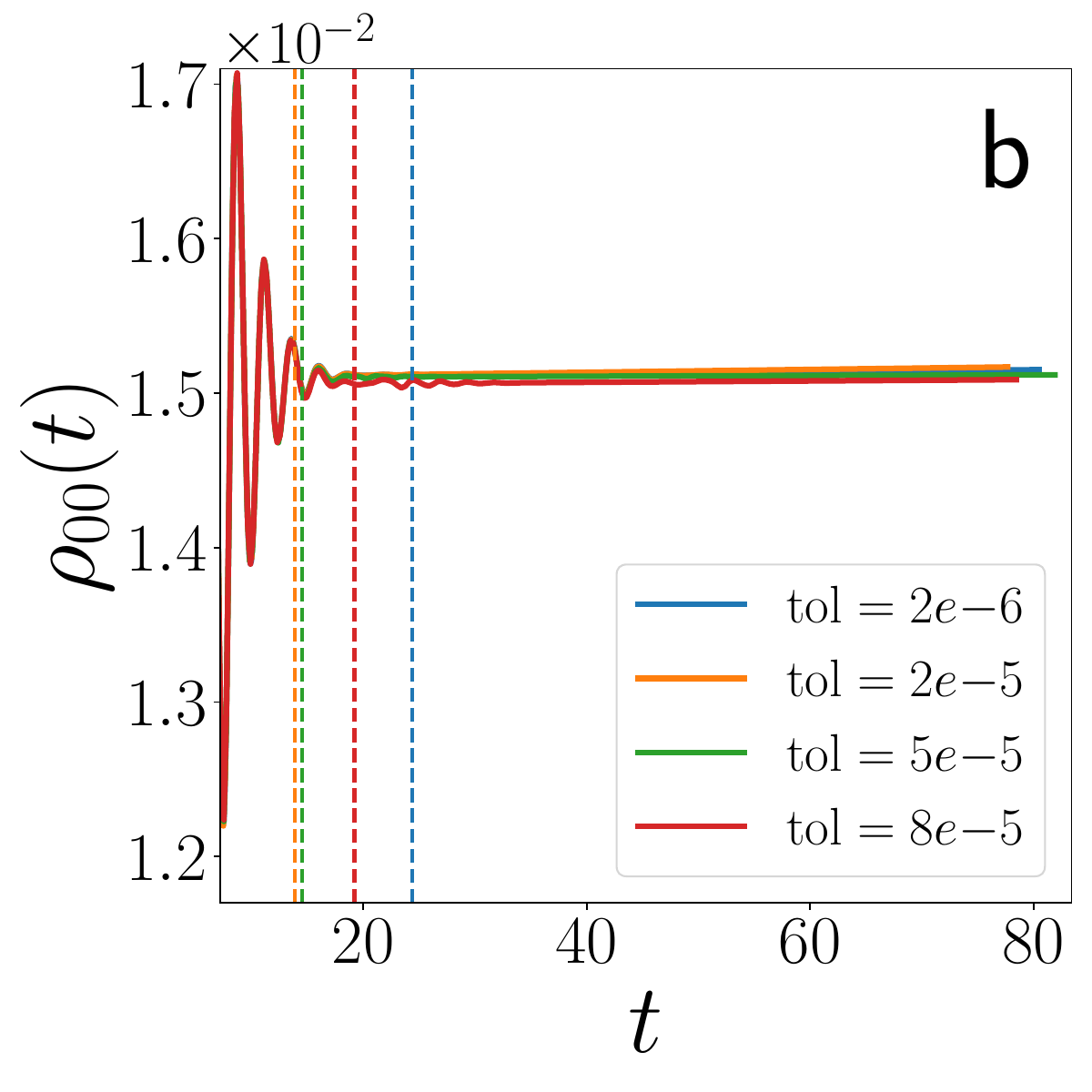}
    \caption{ Instability in the solution of the Kadanoff-Baym Equations induced by the change in step-size when using a fixed history integration order.  Solid lines are the trajectories of the density matrix, and dashed lines are the locations of step-size changes.  There are no external fields or potential ramps applied to this system. The integrator tolerances are atol=rtol for all calculations.  \textit{a:} Using the trapezoid rule for all history integral evaluations. \textit{b:} Using an adaptive integration order with the max interpolation order, $K=20$.}
    \label{fig:instability}
\end{figure}

We first look at the effects of changing the interpolation order within the history integral calculations.  Figure~\ref{fig:instability}(a) shows how the accuracy of the solution is degraded when using a fixed, second order integration method. Solution instabilities manifest as rapid oscillations in the density matrix which are damped after a few oscillations.  The density matrix then develops a constant drift, where it should be constant in the steady-state regime.  Figure~\ref{fig:instability}(b) shows how adapting the interpolation order, with $K=20$, removes these instabilities and drastically improves the accuracy of the solutions at the same tolerance values.

\begin{figure}[htbp!]
    \centering
    \centerline{\includegraphics[width=0.47\textwidth]{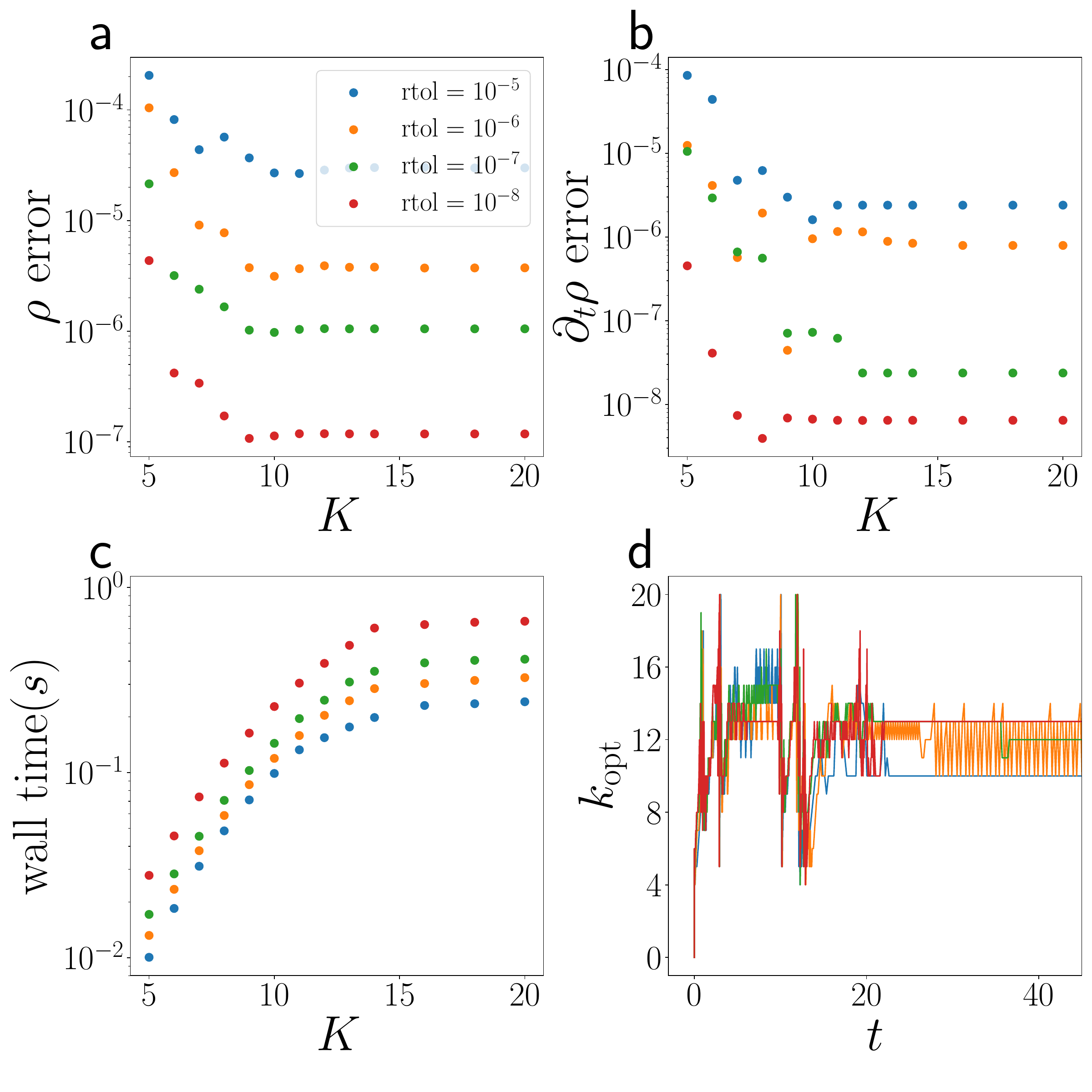}}
    \caption{Analysis of the interpolation scheme used to calculate the integration weights which appear in the evaluation of the history integrals.  The parameter atol is set to $10^{-10}$ for each value of rtol.
    \textit{a:} Relative error in the density matrix for increasing maximum interpolation order. The error bottoms out and becomes constant after order 10 for all values of the relative tolerance (rtol). 
    \textit{b:} The slope of the density matrix in the steady state, which should become zero.  Again the error becomes minimized and constant around order 10.
    \textit{c:} Wall-clock time for the order selection process increases then levels off.  Times are only around 1\% of total run time.  
    \textit{d:} Chosen interpolation orders within calculations at different relative tolerances.  Drops in the order at $t=\{1,3,10,12\}$ correspond to drops in the time step size, where we apply ramps to the system. Colors as in panel \textit{a}.}
    \label{fig:interpolation}
\end{figure}

Figure~\ref{fig:interpolation}(a,b) shows the relative error in the density matrix and its time derivative as the maximum interpolation order, $K$, is increased.  This error is presented for several different relative tolerances to show that the optimization procedure is general for a wide range of tolerances.  The relative error in the density matrix steadily decreases as the interpolation order increases for all tolerances until approximately order $K=10$, after which it stays constant.  In the long-time limit, the system should reach equilibrium, and the density matrix should become constant.  We see that the slope of the density matrix in this steady-state regime also decreases towards zero as we increase the maximum interpolation order.  This decrease shows that these approximations are controlled, and the maximum order can be set as high as desired without impacting solution quality.  Figure~\ref{fig:interpolation}(c) shows that this optimization procedure scales as a power of the max order before leveling off.  These times constitute only around 1\% of the total wall-clock time, which can be seen by comparing to Figure~\ref{fig:CostErr}.  We see the reason for the leveling off of the wall-clock time by looking at the optimal orders in Figure~\ref{fig:interpolation}(d) and noting the interpolation rarely goes above certain orders.  This leveling off of the accuracy and wall time, along with the inexpensive optimization procedure means that the maximum interpolation order can be set to any large value.

\subsection{Accuracy}
\begin{figure}[htbp!]
    \centering
    \centerline{\includegraphics[width=0.47\textwidth]{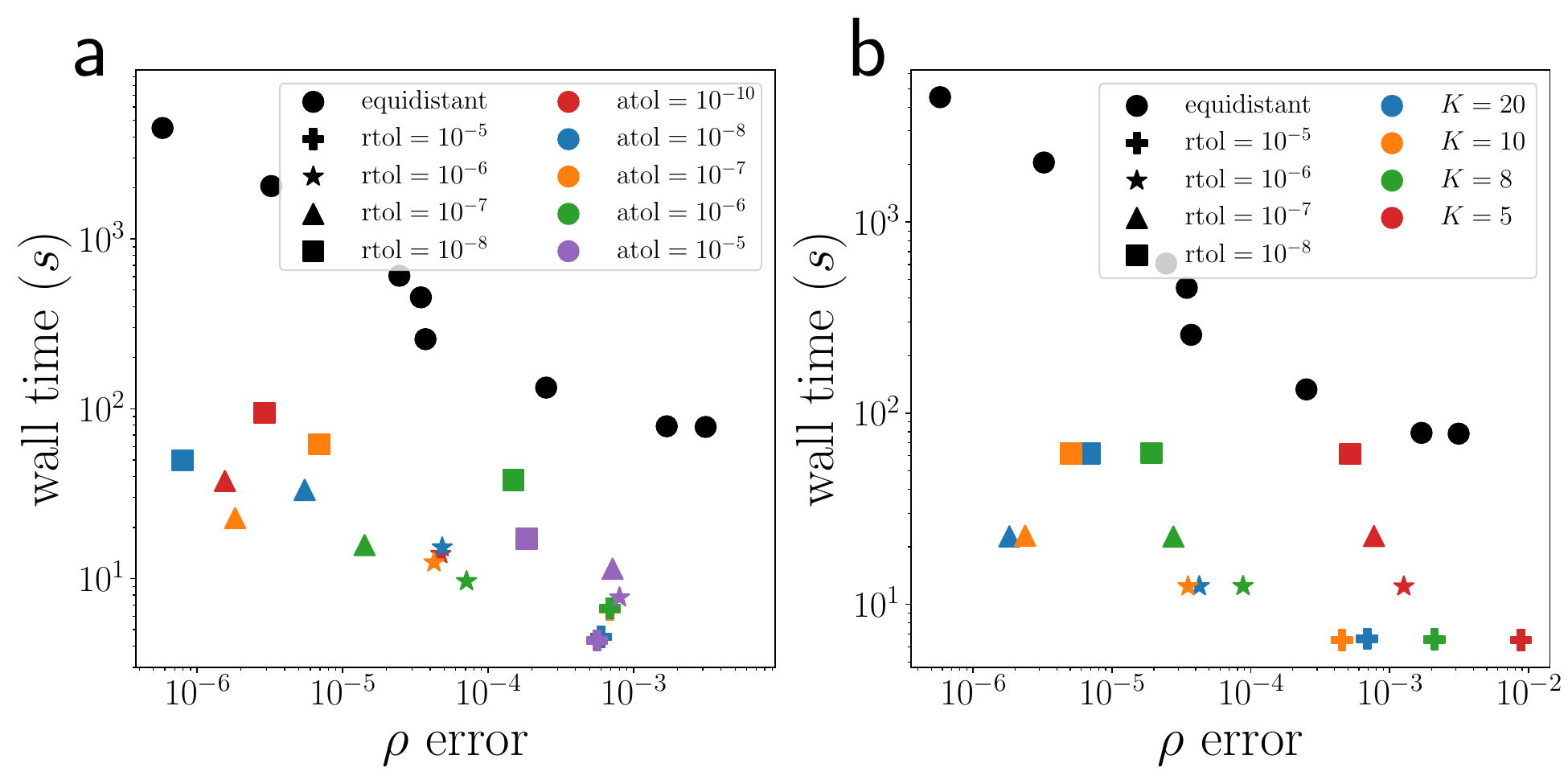}}
    \caption{Relative error of the density matrix compared to the wall-clock time for integrating the system to final time $T_f$. 
    Black circles are for calculations done using the existing uniform step size code.
    \textit{a:} As we decrease both the absolute and relative tolerances, there is a definite trend towards higher accuracy.
    \textit{b:} At a given set of relative and absolute tolerances, the maximum interpolation error plays a large role in increasing accuracy while not affecting the compute cost.}
    \label{fig:CostErr}
\end{figure}

Work-precision data is presented in Figure~\ref{fig:CostErr}, where we compare the relative error in the density matrix against the wall-clock time to reach the final simulation time of $T_f = 81.92$. 
Figure~\ref{fig:CostErr}(a) corresponds to results using relative tolerances ranging from $10^{-5}$ to $10^{-8}$ and absolute tolerances of $10^{-5}$ to $10^{-10}$ with a max history integral order of 20.
It is evident that the error is well controlled by the tolerances and the computational cost compared to the constant step calculations at the same error levels are reduced by an order magnitude.
Figure~\ref{fig:CostErr}(b) also shows results for relative tolerances ranging from $10^{-5}$ to $10^{-8}$ but with an absolute tolerance of $10^{-10}$ and max history integral orders of 5, 8, 10, and 20. 
For a given pair of tolerances, the simulation accuracy improves as the maximum order of the history integral is increased with a negligible impact on the overall simulation time.

\begin{figure*}[htbp!]
    \centering
    \includegraphics[width=\textwidth]{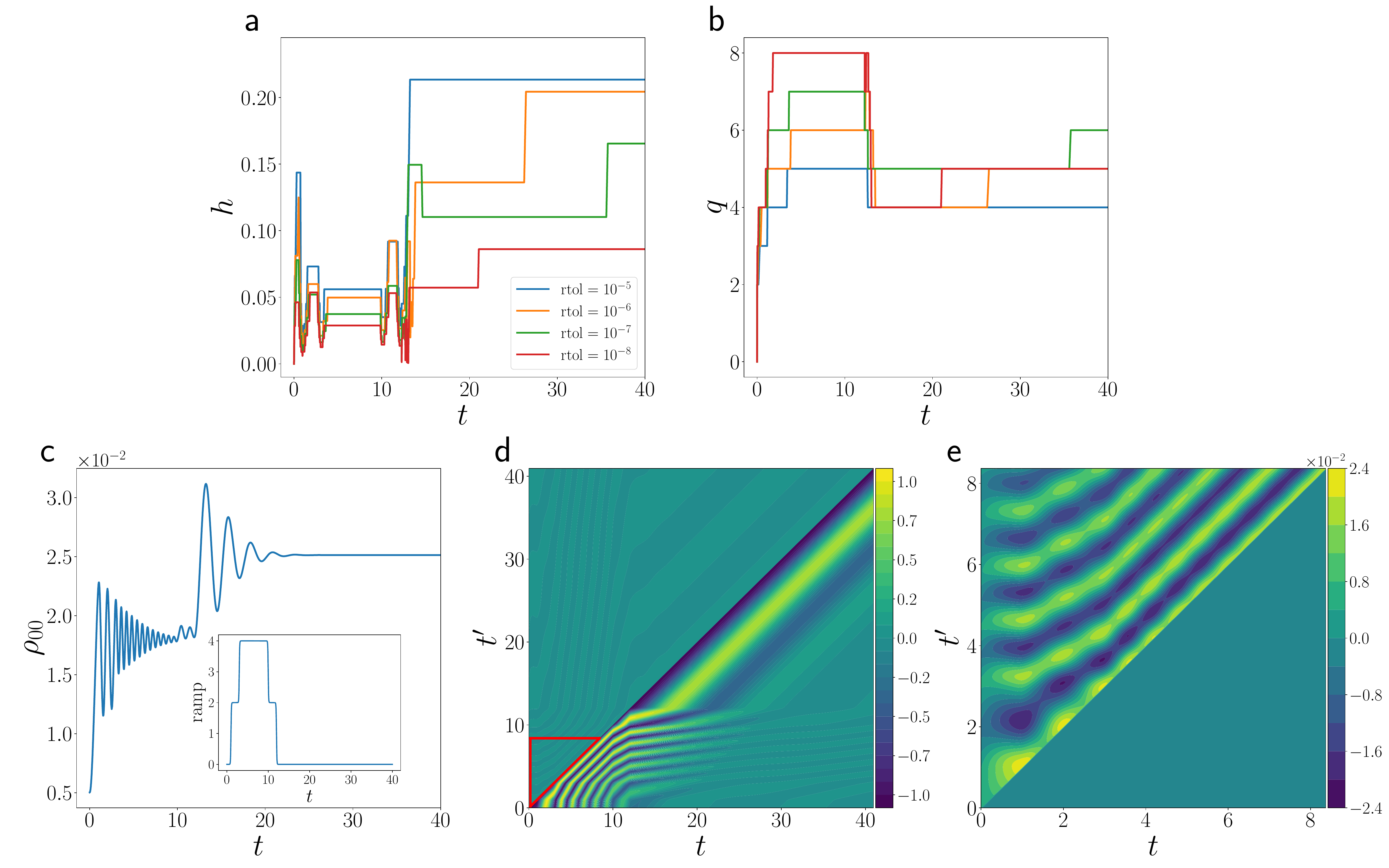}
    \caption{Solutions and time step and order profiles for calculations with four ramps, located at $t=\{1,3,10,12\}$.  
    \textit{a:} Time step profiles for several different tolerances.  
    \textit{b:} Method orders chosen by the adaptive time integrator. 
    \textit{c:} Component of the density matrix corresponding to the site that has a ramp applied.  (Inset: the ramp function we use, Eq.~\ref{Eq:ramp})
    \textit{d:} Imaginary part of the $(0,0)$ component of the lesser (greater) Green's function above (below) the $t=t'$ diagonal. 
    \textit{e:} Zoom on the enclosed red region of \textit{d:} containing the lesser component, which is washed out by the greater component. The solution along the diagonal corresponds to data in \textit{c}.}
    \label{fig:profiles}
\end{figure*}

The source of the efficiency gains are made clear in Figure~\ref{fig:profiles} which shows how the time step sizes and Adams method order adaptively respond to the applied ramp functions. The ramps are located at $t=\{1,3,10,12\}$, which are clearly visible in the time step profiles shown in (a) as the times where the integrator reduces the time step size.  We can see in (c) at those times the density matrix also changes the frequency at which it is oscillating.
When the oscillation slows and eventually stops, the time step sizes increase by large amounts  as the overall dynamics subside.
The integrator also adapts the method order, as shown in (b), to maximize the step size while meeting the requested accuracy. This leads to a rapid increase in order early in time to a steady value with a significant drop around the ramp at $t=12$ followed by a regime of less frequent order changes as the overall dynamics continue to slow.
In (d) and (e) we plot the imaginary part of the ramped component of the Green's functions, where $G^<(t,t')$ is above the diagonal and $G^>(t,t')$ is below the diagonal.  Note the large difference in time scales during and after the ramp.  There are two orders of magnitude difference in the solutions above and below the diagonal in (d), so we plot just the lesser component in (e) to better reveal the structure of the solution.

\subsection{Fixed Point Iteration and Self Consistency}

The expressions for the self-energy in \eqref{Eq:SE} and \eqref{Eq:eps} come from a $\Phi$-derivable theory, which ensures the conservation of several important quantities such as total particle number and energy.  This theory requires that the self-energy must be self-consistent with the Green's functions that are computed from the KBE.  In order to ensure this self-consistency, we use a fixed point iteration scheme within each time step.  

\begin{figure}[htbp!]
    \centering
    \includegraphics[width=0.47\textwidth]{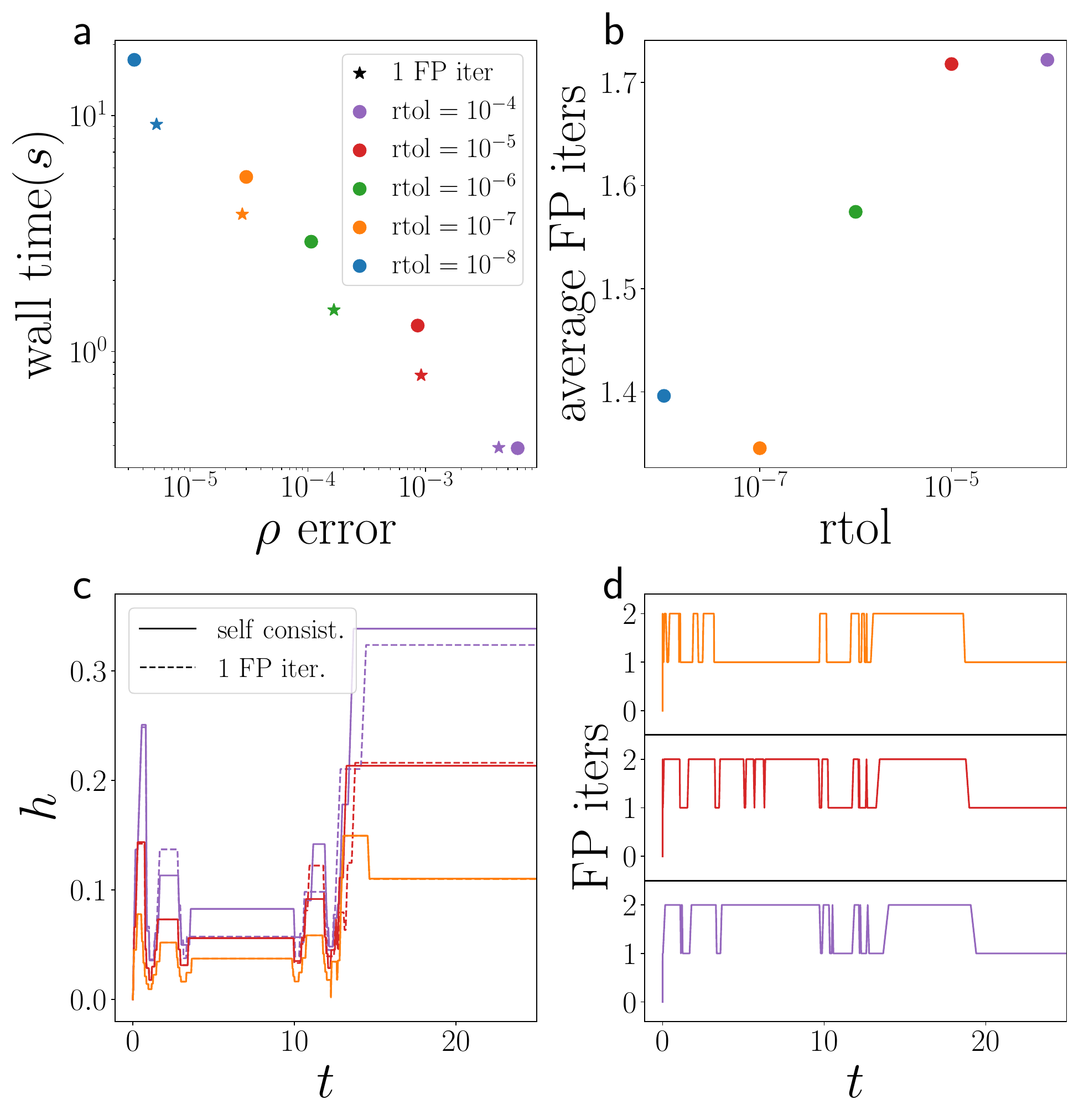}
    \caption{Analysis of the importance of the fixed point iteration.
    \textit{a:} Iterating to convergence leads to a slight increase in accuracy, but with an additional cost associated with reevaluation of    history integrals.
    \textit{b:} The average number of iterations increases with looser tolerances.  
    \textit{c:} Iterating has little impact on the time step sizes, except with the loosest tolerances, where iterating allows for larger step sizes from $t=4$ to $t=10$. Colors as in \textit{a}.  \textit{d:} Number of fixed point iterations taken by CVODE for several different tolerances.}
    \label{fig:NLS}
\end{figure}

Figure~\ref{fig:NLS} presents the results for integration of the KBE when iterating the fixed point solver to convergence as discussed in Section \ref{SSec:ATS} and when applying only a single iteration without a convergence test. 
For relative tolerances ranging from $10^{-8}$ to $10^{-5}$ with an absolute tolerance of $10^{-10}$, the time step profiles are almost identical for the two cases, and the error in the density matrix is slightly decreased when testing for convergence.  
This slight decrease in error comes with increased computational cost, as the fixed point iteration must reevaluate the history integrals for each iteration.  The number of fixed point iterations taken by CVODE are displayed in Figure~\ref{fig:NLS}(d).
We see that the solver takes 2 iterations, up to around time $t\approx20$, which is also the time that the dynamics of the density matrix die off, as seen in Figure~\ref{fig:profiles}(c).  
Once the system enters a steady-state regime, the fixed point iteration is no longer needed, and we take only one iteration to converge.  
We see in Figure~\ref{fig:NLS}(b) that the average number of iterations taken by the solver decreases as we tighten the integrator tolerance.  
This decrease intuitively makes sense, as larger time steps take the system farther away from linearity,
and nonlinearities must be resolved for each step.  
An interesting case is the setup using a relative tolerance of $10^{-4}$. When the iteration is checked for convergence the time step sizes are not as large as when testing for convergence.
The resulting smaller steps lead to an increase in accuracy but a corresponding increase in the computational cost.

\begin{figure*}[htbp!]
    \centering
    \includegraphics[width=\textwidth]{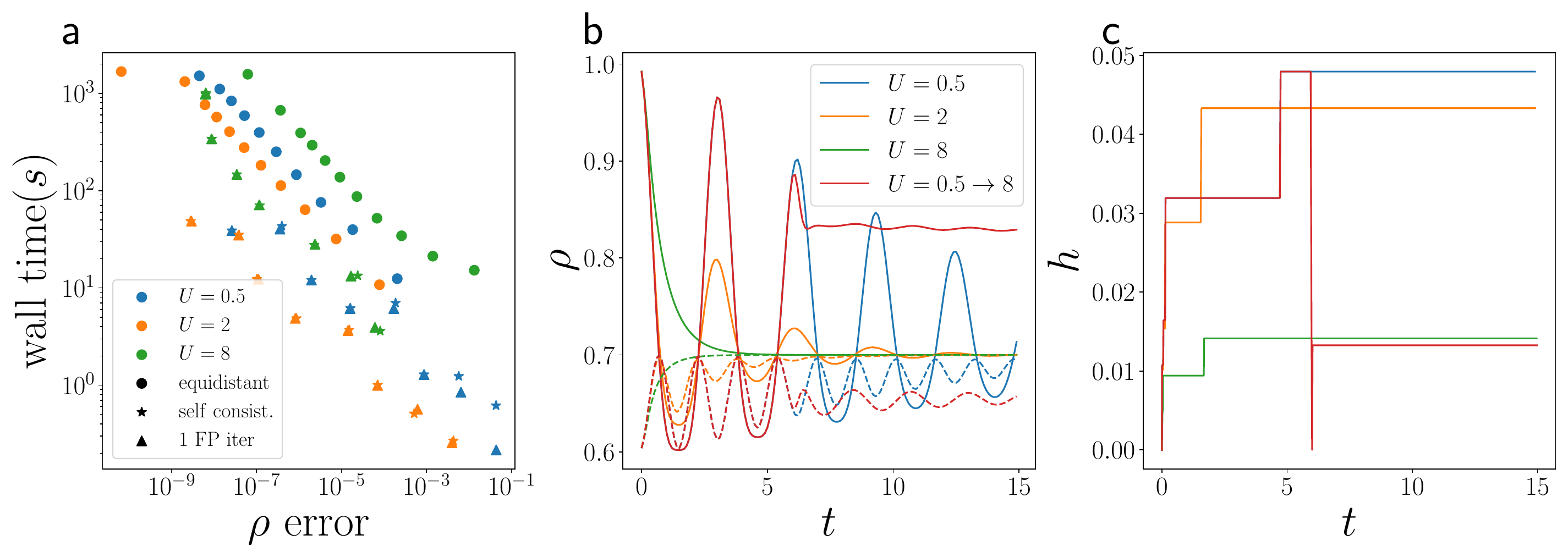}
    \caption{Dynamics of the density matrix for a $2\times2$ Hubbard ring.  Simulation time, $t$, and timestep sizes, $h$, are reported in units of inverse hopping amplitude.  \textit{a:} Wall-time for integration of the system to final time $T_f=15$ compared to the error in the density matrix for the adaptive approach with (stars) and without (triangles) converging the fixed point iteration. The uniform step BDF method (circles) is shown for comparison. Relative tolerances for the adaptive code range from $10^{-3}$ to $10^{-10}$ with an absolute tolerance of $10^{-10}$.  Each point indicates either a different tolerance or time step size. \textit{b:} Dynamics of the density matrix for several values of $U$, as well as a case where $U$ is increased from 0.5 to 8 at $t=6$. Solid and dashed lines correspond to different components of the density matrix. \textit{c:} Time step profiles for the presented density matrix dynamics. Colors as in \textit{b}.}
    \label{fig:Hubbard}
\end{figure*}

\section{Hubbard Physics Results}
\label{Sec:Hubbard_Results}

The Hubbard model \cite{Hubbard1963,SimonsHubbard2015} is the paradigmatic model of interacting electron physics.  
The model has only one free parameter, $U$, the strength of the Coulomb repulsion.  $U$ is reported in units relative to the electron kinetic energy, and times are in the inverse of this unit.  
 The Hubbard model allows us to analyze the importance of the self-consistent fixed-point iteration for a large range of interaction strengths.  To do this, we use the second order GF2 approximation, in which the self-energy is proportional to $U^2$.  For larger interaction strengths we can better understand how the stiffness of the KBE is affected.  Further work is needed in order to utilize the T-matrix or GW approximations~\cite{nessi,TGW}.
We set up a $2\times 2$ Hubbard ring, which leads to a Green's function of size $M=4$.  
The system is prepared out of equilibrium with an excess of electron density on a single site.
We expect the system to evolve towards an equilibrium  state where there is equal density on all sites. Relative tolerances ranging from $10^{-3}$ to $10^{-10}$ are considered with an absolute tolerance of $10^{-10}$.

Results for this problem are shown in Figure~\ref{fig:Hubbard}, where the interaction, $U$, is varied from $0.5$ to $8$.  The value $U=0.5$ is in the weakly interacting regime, and $U=8$ is in the strongly interacting regime \cite{HubbRev_comp,HubbRev_theo}.
Interestingly, we see that even if there is no time-dependence in the Hamiltonian, there is still an order of magnitude advantage in wall-clock time with the adaptive step approach compared to the constant step method at a similar level of accuracy for the density matrix. 
It is also apparent that only a single fixed point iteration is necessary for self-consistency with $U$ values and the range of tolerances considered, as the cost vs. error is nearly identical when taking a single iteration or iterating to convergence.

Figure~\ref{fig:Hubbard}(b) shows the dynamics of two different sites after the non-equilibrium density matrix is allowed to evolve in time.  It is evident that the larger values of $U$ prevent the electrons from moving around much before finding the steady-state.  For weak interactions, the energy cost for double occupation is much lower, and, therefore, the charges oscillate heavily throughout the system as the steady-state is approached.  In order to demonstrate the effectiveness of the adaptive method, we introduce a large change in the system, where the interaction, $U$, is changed on one site from $0.5$ to $8$ at time $t=6$.   
This increased interaction greatly decreases the dynamics of the density matrix, as the amplitude of oscillations become much smaller.  
To resolve this step function change in interaction, the integrator ramps the step-size down very close to zero before recovering to a value very close to the step-size where $U=8$ throughout the entire calculation.  
This behaviour makes sense as the underlying dynamics of the Green's function at the ramped interaction site should closely resemble the dynamics of the $U=8$ calculation.

\section{Thermal Branch Results}
\label{Sec:thermal}

In order to study the effectiveness of the adaptive integration of the KBE with the inclusion of the thermal branch, we implement the model of photoexcited superconductors studied in \cite{blommel2024chirped}.  In this model, an attractive Hubbard interaction facilitates the creation of a superconducting state.  An external electric field is applied to the system, which excites a Higgs mode.  This excitation manifests as an oscillation in the superconducting order parameter, which is the off-diagonal component of the auxiliary density matrix, $\phi = \rho_{01}(t)$. Here we reproduce the results from \cite{blommel2024chirped} with an external electric field intensity of $E_0=0.08$ using the adaptive integration scheme.

\begin{figure*}[htbp!]
    \centering
    \includegraphics[width=0.8\textwidth]{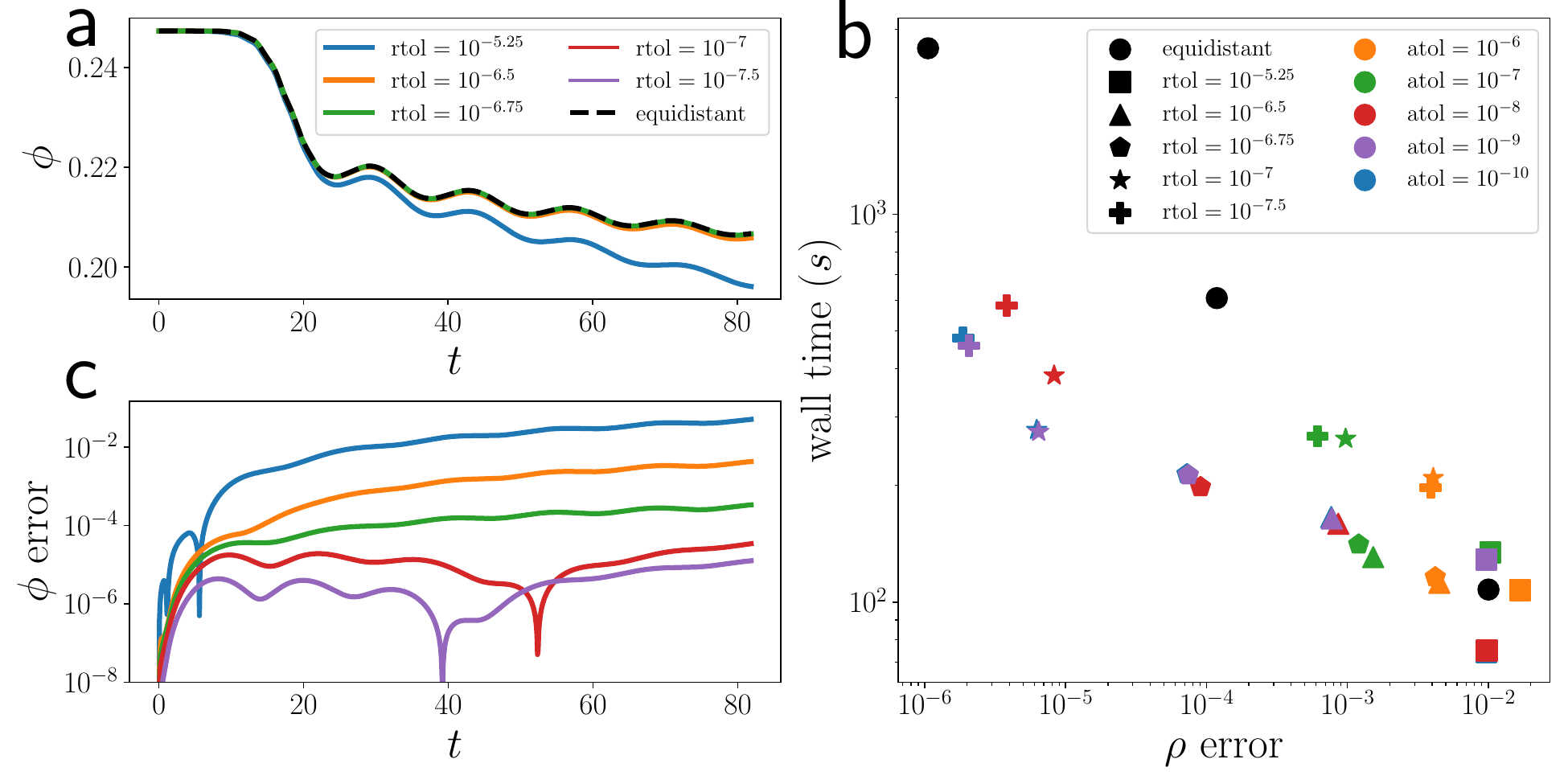}
    \caption{Results for adaptive time stepping of the superconducting system using KBE with the inclusion of a thermal branch.  \textit{a:} Dynamics of the order parameter after excitation from an external electric field for several relative tolerances and an absolute tolerance of $10^{-10}$. \textit{b:} Comparison of error and costs for the adaptive method (colors) and the uniform step BDF method (black).  \textit{c:} Relative difference between the order parameter obtained from the adaptive time stepping method and a self-converged uniform step calculation. Tolerances are the same as in \textit{a}.}
    \label{fig:SC}
\end{figure*}

It should be noted that the results in \cite{blommel2024chirped} are obtained using the hodlr compression algorithm \cite{KayeComp2021}, which is much faster than the cubic scaling of NEDyson and this adaptive implementation.  In order to make the timings comparable, we also implement this model in the NEDyson code.  
Figure~\ref{fig:SC}(a) plots the dynamics of the order parameter for several different relative tolerances and an absolute tolerance of $10^{-10}$ as well as a self-converged uniform step result with $h=0.005$.  Figure~\ref{fig:SC}(c) shows that the relative difference between the adaptive and uniform step codes becomes increasingly small as the tolerances are tightened.  The cost vs. accuracy analysis is presented in Figure~\ref{fig:SC}(b), and shows that the error is systematically decreased as both relative and absolute tolerances are tightened.  In the region of very loose tolerances, the results are comparable to those of the fixed timestep integrator, however in this region the solutions are not acceptably accurate.  As the desired accuracy of the integrator is increased, the advantages of the adaptive integrator become apparent, with around a half an order of magnitude speed up at the same level of accuracy compared to the uniform step results.  

\section{Conclusions}
\label{Sec:Conclusions}

In this work we have presented a time integration scheme for the Kadanoff-Baym equations which is adaptive in both method order and time step size.  This work builds on the integration techniques of \cite{meirinhos2022adaptive} by the inclusion of adaptive order time integration methods, adaptive order for the evaluation of history integration, and self-consistent fixed point iterations.  This adaptive methodology will facilitate the study of quantum dynamics that vary on multiple timescales.  This advantage does not limit our ability to evaluate observables of interest, as Keldysh convolutions necessary for evaluating energy and current, for example, can be done using the optimal integration weights discussed in Sec. \ref{SSec:IW}.  For the evaluation of spectral functions, which involve taking Fourier transforms, the Green's function can be evaluated on a fixed-step temporal grid.

We found that to maintain the accuracy of the integration process with the adaptive step and order schemes presented, the self-energy integrations must include further adaptivity in the order of integrations.  A new scheme that adapts that integration order with a method to choose the optimal order was explored and shown to be highly efficient while maintaining the accuracy of the overall process.

We have applied this method to study several systems that have time-dependent Hamiltonians which vary on time scales much shorter than the time scales of the Green's functions, including on-site potential ramps and interaction quenches.  We have extended the adaptive technique to allow for the integration of quantum systems at a finite temperature and have applied it to a model of photoexcited superconductivity.  For all systems studied, we see around an order of magnitude decrease in computational time for similar accuracy when compared to fixed time step integrators, even in the case where there is no explicit time-dependence of the Hamiltonian.

Due to the non-linear dependence of the self-energy on the Green's function, we use a fixed point iteration to ensure self-consistency conditions.  We have found, however, that for the systems studied, the inclusion of the fixed point iteration does not lead to performance gains when compared to integration without its inclusion.

In the future, this methodology can be extended to further push the maximum obtainable integration times by leveraging methods which exploit the low-rank nature of the Green's function.  Due to the forms of the KBE studied here, it is possible to further improve this method by implementing a parallelized integrator, further allowing for longer times and larger systems to be studied.
\vspace{0.2in}

\section*{Acknowledgements}

We appreciate the valuable discussions with the group of Meirinhos, Kajan, Kroha, and Bode, who have also made their code available through an open source repository.
T.B. (after Aug 2023), E.G., C.W., and D.G. were supported by the U.S. Department of Energy, Office of Science, Office of Advanced Scientific Computing Research and Office of Basic Energy Sciences, Scientific Discovery through Advanced Computing (SciDAC) program with T.B. and E.G. specifically under Award No. SC0022088.  
T.B. was supported by the U.S. Department of Energy, Office of Science, Office of Advanced Scientific Computing Research, Department of Energy Computational Science Graduate Fellowship under Award Number DE-SC0020347 until Aug. 2023.
This work was performed in part under the auspices of the U.S. Department of Energy by Lawrence Livermore National Laboratory under Contract DE-AC52-07NA27344. LLNL-JRNL-863703.

\bibliography{mybib}

\end{document}